\documentclass{aa}
\usepackage{graphicx}
\usepackage{calc}
\usepackage{natbib}
\bibpunct{(}{)}{;}{a}{}{,}

\begin{document}
\title{\object{Palomar~13}: a velocity dispersion inflated by binaries~? \thanks{Tables~\ref{tableallstars} and \ref{tablemembership} are also available in electronic form
at the CDS via anonymous ftp to cdsarc.u-strasbg.fr (130.79.128.5)
or via http://cdsweb.u-strasbg.fr/cgi-bin/qcat?J/A+A/}}

   \author{A. Blecha
          \inst{1}
          \and
          G. Meylan\inst{2}\fnmsep\thanks{
          Affiliated  with the Space Telescope Division of 
          the European Space Agency, ESTEC, Noordwijk, Netherlands}
          \and
          P. North
          \inst{3}
          \and
          F. Royer
          \inst{1}
          }

   \offprints{A. Blecha}

   \institute{Observatoire de Gen\`eve, Universit\'e de Gen\`eve, 
              Chemin des Maillettes 51, 1290 Sauverny, 
              Switzerland \\
              \email{Andre.Blecha@obs.unige.ch, Frederic.Royer@obs.unige.ch}
         \and
              Space Telescope Science Institute,
              3700 San Martin Drive, Baltimore, MD 21218, U.S.A.\\
              \email{gmeylan@stsci.edu}
         \and
              Laboratoire d'Astrophysique, 
              \'Ecole Polytechnique F\'ed\'erale de Lausanne, 
              Observatoire, 1290 Chavannes-des-Bois, 
              Switzerland \\
              \email{Pierre.North@obs.epfl.ch}
             }

   \date{Received September 15, 2003; accepted September 16, 2003}

\titlerunning{Pal~13: a velocity dispersion inflated by binaries~?}


   \abstract{

Recently, combining radial  velocities from Keck/HIRES \'echelle spectra
with published proper motion membership probabilities, \citet{Coe_02} 
observed a sample of $21$ stars, probable members of Palomar~13,
a  globular cluster in  the Galactic  halo.  Their  projected velocity
dispersion $\sigma_\mathrm{p}  = 2.2 \pm 0.4$~km\thinspace s$^{-1}$\  gives a mass-to-light
ratio  $\mathcal{M}/L_V =  40^{+24}_{-17}$,  about one  order  of
magnitude larger than the usual estimate for globular clusters.

We present  here radial velocities  measured from three  different CCD
frames  of commissioning  observations obtained  with the  new ESO/VLT
instrument  FLAMES  (Fibre Large  Array  Multi Element  Spectrograph).
From these  data, now publicly  available, we measure  the homogeneous
radial  velocities   of  eight  probable  members   of  this  globular
cluster. A new projected velocity dispersion $\sigma_\mathrm{p} = 0.6$--$0.9 \pm
0.3$~km\thinspace s$^{-1}$\ implies  Palomar~13 mass-to-light ratio $\mathcal{M}/L_V =  3$--$7$,  similar to the  usual value for globular  clusters.  We
discuss briefly the two most  obvious reasons for the previous unusual
mass-to-light ratio finding: binaries,  now clearly detected, and more
homogeneous data from the multi-fibre FLAMES spectrograph.

   }

   \maketitle


\section{Introduction}

All the dynamical studies of nearby globular clusters have established
that these  dynamical systems contain  no dark matter, apart  from the
expected  stellar remnants  such  as white  dwarves  and neutron  stars
\citep[e.g.,][]{PrrMen93,MenHee97}.  Consequently,
globular clusters may be the  most massive stellar systems in which no
non-baryonic  dark  matter is  dynamically  detected, while  dynamical
evidence  for non-baryonic  dark matter  seems to  be present  in most
galaxies, from the  faintest dwarf spheroidals (dSphs) to the brightest
cDs galaxies, and clusters of galaxies as well.

Some  of  the  local  dSphs, around the Galaxy and M31, have integrated absolute
luminosities  similar  or  fainter than those of the brightest Galactic globular
clusters.  Since  there  is  evidence  that  some  Galactic  dSph  galaxies  are
dark-matter-dominated,  it  is  therefore  reasonable  to check if some globular
clusters  do  present  dynamical  evidence  for non-baryonic dark matter. Such a
possibility  may  be  emphasized by the current predictions of Cold Dark Matter
(CDM) numerical simulations of galaxy formation, in which the number of low-mass
dark-matter  substructures orbiting the halo of massive galaxies largely exceeds
the  number  of  dwarf galaxies observed in the halos of both our Galaxy and M31
\citep[e.g., ][]{Kln_99,Moe_99,Moe_01}. However, recently
improved CDM models may correctly predict the observed number of satellite galaxies
\citep{Biy03}.

Could some of the globular clusters in the outer parts of the Galactic
halo be such dark-matter  substructure~?  These remote stellar systems
have so  far been  poorly studied because  of the difficulties  in the
acquisition  of high-quality  radial velocities  and  proper motions,
direct  consequences of the  faintness and  sparsity of  these distant stellar
systems. They  are nevertheless important probes of  the formation and
evolution  of the  Galaxy,  as their  ages  and metallicities  provide
direct constraints  on the duration  of halo formation process  and on
the time-scale  for Galactic chemical enrichment, while  the shape and
extent  of the  Galactic dark  halo are  constrained by  their orbital
properties.

In  1998, a  program  was started  at  Californian Institute of Technology to  study the  internal
dynamics  of  seven distant  halo  globular  clusters  using the  High
Resolution   Echelle    Spectrometer   (HIRES)   at    the   W.~M.~Keck
Observatory. The aim was the first direct measurements of the velocity
dispersions and mass-to-light ratios for these clusters.  Six clusters
in this sample exhibited  velocity dispersions $\sigma_\mathrm{p} \sim 1$~km\thinspace s$^{-1}$,
translating  into  mass-to-light  ratio  values  typical  of  globular
clusters $\mathcal{M}/L_V\sim 3$ (all mass-to-light ratios quoted
in this paper are in solar units). Only one cluster, the halo globular
cluster  Palomar~13,  displayed   a  velocity  dispersion  larger  than
expected.

\citet[ also referenced  below as  the  Keck study]{Coe_02} 
presented  a  careful   analysis,  combining  radial  velocities  from
Keck/HIRES \'echelle spectra  with published  proper  motion membership
probabilities from \citet{Sil_01}.  They obtained a sample of 21
stars,  probable members  of Palomar~13.   Their  projected, intrinsic
velocity  dispersion  of  $\sigma_\mathrm{p}   =  2.2\pm 0.4$~km\thinspace s$^{-1}$\  implied  a
mass-to-light ratio $\mathcal{M}/L_V= 40^{+24}_{-17}$, about one
order of magnitude larger than  the usual value for globular clusters.
C\^ot\'e et al.\ discussed at length all  possible reasons for
such an  unusual result:  (i) some velocity  ``jitter'' among  the red
giants, (ii) a  few binary stars, (iii) a  non-standard mass function,
(iv) process of dissolving into the Galactic halo through catastrophic
tidal  heating during  a  recent perigalacticon  passage,  or (v)  the
presence of a massive non-baryonic dark matter halo.

It  is worth  emphasizing that,  in C\^ot\'e et al., careful
determination  of the error  bars made  the usual  mass-to-light ratio
values for globular clusters $\mathcal{M}/L_V\sim 3$ at about two
sigmas from the $\mathcal{M}/L_V$ value obtained for Palomar~13.

Because  of  this  marginally  significant  and  puzzling  result  from
C\^ot\'e et al., some more  spectroscopic data of stars in the
field  of Palomar~13  were acquired  during the  commissioning  of the
ESO/VLT instrument FLAMES.  We  present hereafter new high-quality and
homogeneous radial velocities for 46 stars, 9 of them being members
of Palomar~13, which provide new velocity dispersion and mass-to-light
ratio values for  this globular cluster.  The remaining  of this paper
is  as  follows:  Sect.~2  presents  the  observation  and  the  data
reduction, Sect.~3 discusses the membership of the stars, Sect.~4
gives  the  new velocity  dispersion  and corresponding  mass-to-light
ratio, and Sect.~5 discusses the plausible reasons for the difference between the
present results and those obtained by C\^ot\'e et al.


\section{Observations and data reduction}

\subsection{Observations}

All  the new data  presented in  this paper  were acquired  between
August 28 and  September 3 2002, during the commissioning  of the ESO/VLT
instrument FLAMES  (Fibre Large Array Multi  Element Spectrograph), at
Paranal, Chile.  For more information  on this instrument, see the ESO
web  page  {\em http://www.eso.org}  and  the  recent publications  by
\citet{Pai_02} and \citet{Ror_02}.  In order to clarify
some possibly confusing uses  of denominations, it is worth mentioning
that FLAMES is the name of  the instrument, while GIRAFFE is the name of the
spectrograph within FLAMES.

We were  involved neither  in the  selection of the  stars nor  in the
preparation  and acquisition  of these  observations.   Recently, these
data were made publicly available  from the ESO/VLT archives site\footnote{\em
http://www.eso.org/science/flames\_comm/\-FLAMES\_comm\_PAL\_13.html}.

The      Palomar~13      observations      are      summarized      in
Table~\ref{tablelogbook}.   Three  exposures,  each  of  one  hour 
integration, were acquired in early September 2002.  Calibration files
were  acquired  a  few days  earlier.   All  data  were taken  in  the
high-resolution spectrograph setup  HR-9 with spectral domain spanning
from  $514.3$ to  $535.4$  nm and  resolution $\lambda/\delta\lambda  \sim
20\,000$.  Both  positioning plates  of MEDUSA were  used with  the 133
fibres assigned  as follows: 55 fibres  to program stars,  14 to sky
measurements,  5  to simultaneous  wavelength  calibrations, while  59
fibres  remained  unused.  Average  signal-to-noise ratio (SNR) for star with $V=18$ is given as
indication  of  spectra  exposure.  Note that expected accuracy for HR-9 setup and $\mathrm{SNR}=10$ is
$\sigma_{\ensuremath{V_\mathrm{r}}}=0.20$ km\thinspace s$^{-1}$\ \citep[see ][]{Ror_02}.

\begin{table*}[!htp]
\caption{Log-book  of the observations  from FLAMES  commissioning data
archives: Palomar~13 observations and corresponding calibrations
collected in HR-9 setup ($514.3$--$535.4$~nm). SNR is the average
signal-to-noise ratio for star of $V$-$\mathrm{magnitude}=18$.}
\begin{tabular}{clcccccc}
\hline
\hline
\# &Archives file name&Date of & Time &Instrument&Exp. time&SNR & Remark \\
   &              &observations& [UT] &configuration& [s] &at $V=18$ &       \\
\hline 
1&FLAMES\_GIRAF\_OBS244\_0015 &2002-09-01&05:29&Medusa2&3600 & 10 & good\\
2&FLAMES\_GIRAF\_OBS244\_0016 &2002-09-01&06:45&Medusa1&3600 & 10 & acceptable\\
3&FLAMES\_GIRAF\_OBS246\_0003 &2002-09-03&03:41&Medusa2&3600 &  7 & poor\\
\hline
4&FLAMES\_GIRAF\_FLAT240\_0008&2002-08-28&19:51&Medusa1& 143 &flat\\
5&FLAMES\_GIRAF\_WAVE240\_0011&2002-08-28&20:10&Medusa1& 484 &ThAr\\
6&FLAMES\_GIRAF\_FLAT240\_0020&2002-08-28&23:53&Medusa2& 143 &flat\\
7&FLAMES\_GIRAF\_WAVE238\_0018&2002-08-26&19:33&Medusa2& 482 &ThAr\\
\hline
\end{tabular}
\label{tablelogbook}
\end{table*}

\subsection{Data reduction and calibration}

For  each positioning  plate, two  calibration frames  were  used.  The
first  one  --  the flat  field  --  gave  the master  localization  and
analytical model  of the PSF perpendicular to  the spectral dispersion
direction, while the second one  -- ThAr wavelength calibration frame --
provided two pieces necessary  for the master wavelength solution: the
2-D optical solution and the 2-D Chebyshev polynomial correction.  For
the HR-9 setup, approximately $6\,000$ line positions were used to adjust
the 30 parameters  of the solution.  Finally, the  rebinned extracted ThAr
spectra  were cross-correlated  with the  ThAr mask  and  the measured
mismatch between spectra was used to update the slit geometry model.

The spectra were extracted using  the standard {\em Python} version of
BLDRS~-~Baseline   Data    Reduction   Software   (girbldrs-1.09   and
girbldrs-pipe-1.05)            available           from           {\em
http://girbldrs.sourceforge.net}. Basic description  of BLDRS is given
in \citet{Bla_00}.

The most important calibration feature of GIRAFFE is the presence of five
Simultaneous  Wavelength  Calibration (SIMCAL)  ThAr  spectra in  each
exposure.   The SIMCAL  spectra  are optimally  exposed and  regularly
spaced over the detector (spectra \#1, 32, 63, 94, and 125 out of 133
in the present setup).  They  are used to accurately adjust the master
localization and the master wavelength calibration.

The  raw  images were  processed  through  the  following steps:  bias
subtraction, localization adjustment, optimal extraction, rebinning to
linear  wavelength  space  with  step  of $0.005$~nm  using  the  master
wavelength solution, wavelength  solution translation using the SIMCAL
spectra cross-correlation with ThAr mask and the final rebinning using
the translated wavelength solution. Note that we did not subtract dark
(negligible effects) and we  did not flat-field (irrelevant for the
cross-correlation),  the flat-field  frame  being only  used to  derive
master localization  and PSF  model. It is  worth mentioning  that the
final translation  of the master  wavelength solution, which  is below
$0.001$~nm  for all  three science  CCD frames,  indicates  an excellent
instrument stability,  at least over  the period of one  week spanning
the above acquisition of data.

\subsection{Radial velocities}

Radial   velocities    \ensuremath{V_\mathrm{r}}\   were   then    measured   through   the
cross-correlation of the above  spectra with the standard GIRAFFE mask
for the HR-9 setup.  It is  a binary CORAVEL-type mask for the stellar
spectral  type F0,  with  variable line-width  ranging  from $0.005$  to
$0.06$~nm and  a total number of  98 spectral lines. In  this paper, the
new FLAMES data are labeled with  {\em F} for FLAMES while the data
from C\^ot\'e et al.\  are    labeled    with   {\em    K}   for    Keck.
Table~\ref{tableallstars}  summarizes  our  results for  all  observed
stars.  Columns  (1) and (2)  give the star identification  number and
its cluster  membership probability, first from this  FLAMES study and
second from the  Keck study, whenever available.  All  our stellar IDs
come from  \citet{Sil_01}.  Columns (3)  through (6)  give the
equatorial coordinates $\alpha$ and $\delta$, the distance $R$ of the star
to  the   cluster  center  
($\alpha_\mathrm{J2000} = 23^\mathrm{h}\,06^\mathrm{m}\,44\fs 480$, 
 $\delta_\mathrm{J2000} = 12\degr\,46\arcmin\,19\farcs 20$) and the $V$ magnitude from \citet{Sil_01}.
Columns  (7)   through  (9)  give  the   radial  velocities,  whenever
available, from each of our three FLAMES epochs. Column (10) gives for
each star the number $n$ of valid measurements available from  our three
epochs, (11) indicates the  cross-correlation-peak quality $Q$ (1 for best
and 9 for  worst), where the first number  represents the certainty of
the peak  identification and  the second the  quality of the  fit, and
(12) gives the  standard deviation of  the two or  three measurements,
whenever  applicable.  Column  (13) gives  the FLAMES  mean
radial  velocity  from  our $n$ measurements and the estimated error ($\sigma_{\ensuremath{V_\mathrm{r}}}$ or
individual error) and Column  (13) is the  Keck  mean  radial
velocity from C\^ot\'e et al. for comparison.

The  raw  measurements  of  55  stars were  cleaned  by  removing  all
measurements with  unclear peak identification or very  bad quality of
the  fit.  From originally  55 objects,  46 remain  with at  least one
valid  radial velocity  measurement.   We get  4  objects with  all three
measurements (\#26),  30 objects with two measurements  and 12 objects
with one measurement (6 from first plate, 5 from second and 1 from
third plate).

Ignoring the three \ensuremath{V_\mathrm{r}} variable stars \#36 \#38 and \#156 (see discussion
below), we  have an  immediate \ensuremath{V_\mathrm{r}} quality  check for  the $27+4=31$
stars with  at least  two measurements: the  mean of  their individual
standard  deviations   is  $0.54$~km\thinspace s$^{-1}$.   

We  also   include  in  Table~\ref{tableallstars} the 13 stars with one measurement only but having
acceptable cross-correlation-peak  quality $Q$  (Column 11).  Note
that  measurements with  $Q$  values higher  than  2--3 should  be
considered with caution.

\begin{figure*}[!htp]
\centering
\includegraphics[width=12cm]{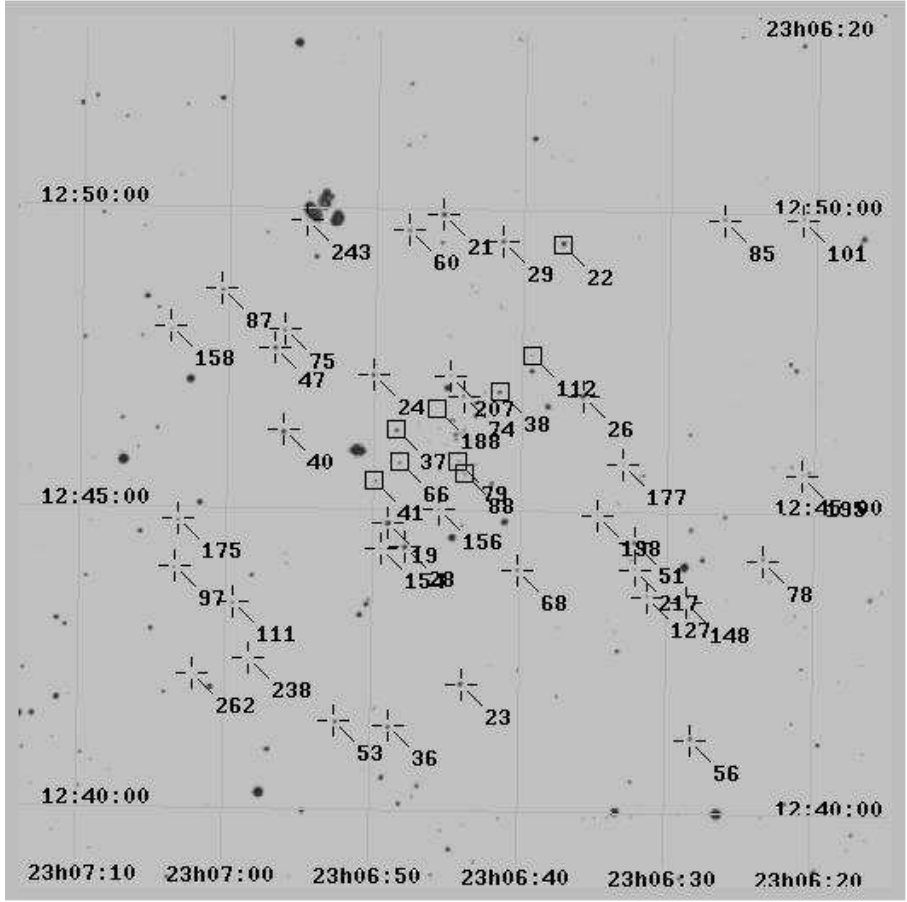}
\caption{ Digital Sky Survey image  with all 46 stars listed in Table~\ref{tableallstars}.   Stars are numbered  according to \citet{Sil_01}.
 Member  stars  are enclosed by boxes and other program stars indicated with crosshairs.
The  entire
FLAMES field (diameter 26\,\arcmin) is shown. See Fig.~\ref{Fig2} for a zoom into the cluster center.}
\label{Fig1}%
\end{figure*}

\begin{figure*}[!htp]
\centering
\includegraphics[width=16cm]{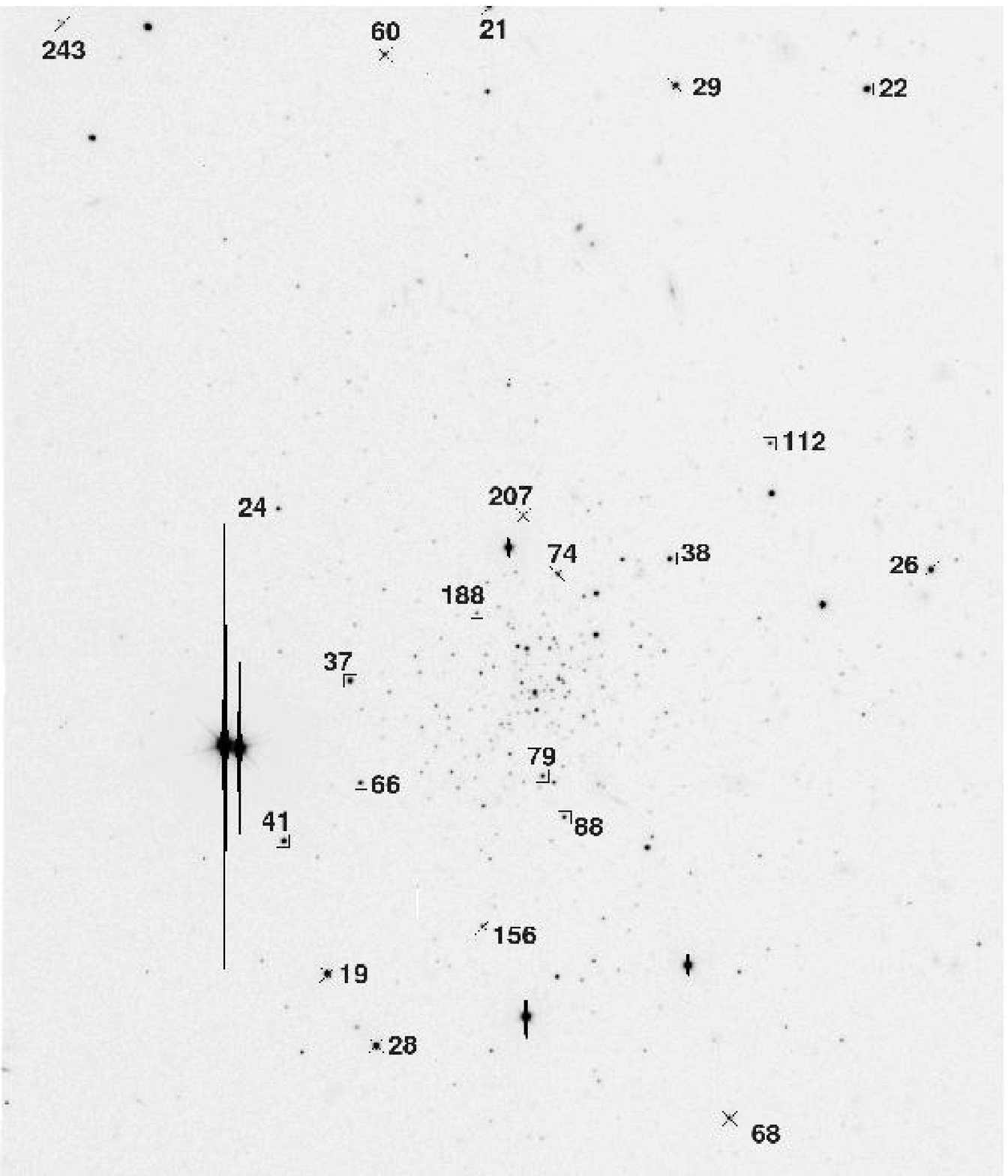}
\caption{Detailed finding chart of the central field: $V$-band
image of Palomar~13 taken with the Low-Resolution Imaging
Spectrometer  on the Keck  II telescope  from \citet{Coe_02}.
Member  stars  are enclosed by boxes and other program stars indicated with crosses.}
\label{Fig2}%
\end{figure*}

Figures~\ref{Fig1} and~\ref{Fig2} give the finding  charts of  all 46  stars measured
with FLAMES.   In Table~\ref{tableallstars},  all stars are  sorted by
increasing  values of their  mean radial  velocities listed  in Column
(13).  In this table, the intermediate horizontal line isolates the 11
stars (at  the bottom of Table~\ref{tableallstars})  with very similar
radial  velocities: they  may be  considered as  cluster  members when
using the radial-velocity membership criterion only (see below).

It  may be  worth  mentioning  that all  radial  velocities listed  in
Columns (7), (8),  and (9) are raw velocities,  obtained directly from
the cross-correlation process, corrected only for the solar system and
earth velocities.  No other attempt  is made to correct for any effect
due  to  various  dependence  on  stellar  spectral  types  (the  same
cross-correlation mask was used for  all spectra) or any effect due to
the limited  spectral range.  The undergoing analysis  of other FLAMES
data sets  not related to  Palomar~13 indicates that the  random error
due to the mask mismatch could be a few $0.1$~km\thinspace s$^{-1}$ while the systematic
shifts due to the limited  wavelength coverage do not exceed a similar
value (a few $0.1$~km\thinspace s$^{-1}$) while using the HR-9 setup.


\begin{table*}[!htp]
\caption{Barycentric radial  velocities for  46  stars  in the  field
centered on Palomar~13.  The extra horizontal line isolates a clump of
11  stars with  radial velocities  very close  to the  systemic radial
velocity of the star cluster, potentially members of the cluster.}
\setlength{\tabcolsep}{3.5pt}
\begin{tabular}{ccrrrrrrrcrrrr}
\hline
\hline
\multicolumn{1}{c}{ID}  &  \multicolumn{1}{c}{M}  &  \multicolumn{1}{c}{$\alpha_\mathrm{J2000}$} &
\multicolumn{1}{c}{$\delta_\mathrm{J2000}$}  &  \multicolumn{1}{c}{$R$}  & \multicolumn{1}{c}{$V$} &
\multicolumn{1}{c}{$\ensuremath{V_\mathrm{r}} 1$}         &         \multicolumn{1}{c}{$\ensuremath{V_\mathrm{r}} 2$}        &
\multicolumn{1}{c}{$\ensuremath{V_\mathrm{r}} 3$}  &  \multicolumn{1}{c}{$n$}  &  \multicolumn{1}{c}{$Q$} &
\multicolumn{1}{c}{$\sigma_{\ensuremath{V_\mathrm{r}}}$}     &     \multicolumn{1}{c}{    $\ensuremath{V_\mathrm{r}}$FLAMES}    & \multicolumn{1}{c}{ $\ensuremath{V_\mathrm{r}}$Keck} \\
\multicolumn{1}{c}{\em F--K} & \multicolumn{1}{c}{\em F--K} &
\multicolumn{1}{c}{[\degr]} & \multicolumn{1}{c}{[\degr]} & \multicolumn{1}{c}{[\arcsec]} & \multicolumn{1}{c}{[mag]} & \multicolumn{1}{c}{[km\thinspace s$^{-1}$]}   & \multicolumn{1}{c}{[km\thinspace s$^{-1}$]}   & \multicolumn{1}{c}{[km\thinspace s$^{-1}$]}   &  \multicolumn{1}{c}{}    &  \multicolumn{1}{c}{}    & \multicolumn{1}{c}{[km\thinspace s$^{-1}$]}           & \multicolumn{1}{c}{[km\thinspace s$^{-1}$]}      & \multicolumn{1}{c}{[km\thinspace s$^{-1}$]} \\
 \multicolumn{1}{c}{(1)}& \multicolumn{1}{c}{(2)} & \multicolumn{1}{c}{(3)}       & \multicolumn{1}{c}{(4)}       & \multicolumn{1}{c}{(5)}                  & \multicolumn{1}{c}{(6)}   & \multicolumn{1}{c}{(7)}    & \multicolumn{1}{c}{(8)}    & \multicolumn{1}{c}{(9)}    & \multicolumn{1}{c}{(10)} & \multicolumn{1}{c}{(11)} & \multicolumn{1}{c}{(12)}           & \multicolumn{1}{c}{(13)}      & \multicolumn{1}{c}{(14)}   \\
\hline
      29 &    0 & $346.67271$ & $12.82458$ & $190$ & $17.33$ & $-248.39$ & $-248.12$ &     ...  & 2 & 11 &  0.19 & $-248.25 \pm 0.19$ &     ... \\
      21 &    0 & $346.68958$ & $12.83175$ & $210$ & $16.91$ & $-153.00$ & $-154.25$ &     ...  & 2 & 11 &  0.88 & $-153.62 \pm 0.88$ &     ... \\
     111 &    0 & $346.74825$ & $12.72392$ & $282$ & $19.14$ &      ...  & $-109.18$ &     ...  & 1 & 12 &   ... & $-109.18 \pm 1.27$ &     ... \\
     101 &    0 & $346.58725$ & $12.83158$ & $410$ & $19.13$ & $ -66.55$ &  	...  & $-67.13$ & 2 & 13 &  0.41 & $ -66.84 \pm 0.41$ &     ... \\
     195 &    0 & $346.58687$ & $12.76064$ & $356$ & $19.85$ &      ...  &  	...  & $-61.50$ & 1 & 23 &   ... & $ -61.50 \pm 1.31$ &     ... \\
     175 &    0 & $346.76379$ & $12.74694$ & $295$ & $19.47$ & $ -57.93$ &  	...  &     ...  & 1 & 23 &   ... & $ -57.93 \pm 2.01$ &     ... \\
 26--103 & 0--0 & $346.64942$ & $12.78197$ & $133$ & $17.05$ & $ -57.38$ & $ -57.19$ & $-58.54$ & 3 & 11 &  0.73 & $ -57.70 \pm 0.73$ & $-56.72$\\
      78 &    0 & $346.59775$ & $12.73714$ & $338$ & $18.66$ & $ -54.43$ & $	...$ & $-53.68$ & 2 & 12 &  0.53 & $ -54.05 \pm 0.53$ &     ... \\
      40 &    0 & $346.73433$ & $12.77156$ & $176$ & $17.27$ & $ -45.06$ & $ -45.47$ &     ...  & 2 & 11 &  0.28 & $ -45.26 \pm 0.28$ &     ... \\
      56 &    0 & $346.61792$ & $12.68717$ & $384$ & $18.21$ & $ -42.36$ & $	...$ & $-42.88$ & 2 & 11 &  0.36 & $ -42.62 \pm 0.36$ &     ... \\
  19--13 & 0--0 & $346.70421$ & $12.74631$ & $112$ & $16.54$ & $ -37.12$ & $ -38.32$ &     ...  & 2 & 11 &  0.85 & $ -37.72 \pm 0.85$ & $-38.90$\\
   68--1 & 0--0 & $346.66737$ & $12.73358$ & $149$ & $18.57$ & $ -36.66$ & $ -37.61$ &     ...  & 2 & 12 &  0.67 & $ -37.13 \pm 0.67$ & $-37.48$\\
 74--101 & 0--0 & $346.68329$ & $12.78169$ & $ 34$ & $18.64$ & $ -18.23$ & $ -17.71$ &     ...  & 2 & 11 &  0.36 & $ -17.97 \pm 0.36$ & $-17.39$\\
      51 &    0 & $346.63421$ & $12.74147$ & $212$ & $18.12$ & $ -16.48$ &  	...  & $-17.05$ & 2 & 11 &  0.41 & $ -16.77 \pm 0.41$ &     ... \\
     177 &    0 & $346.63758$ & $12.76325$ & $174$ & $19.61$ & $ -11.92$ &  	...  & $-10.80$ & 2 & 12 &  0.80 & $ -11.36 \pm 0.80$ &     ... \\
      47 &    0 & $346.73692$ & $12.79464$ & $202$ & $17.90$ &      ...  & $ -10.12$ &     ...  & 1 & 12 &   ... & $ -10.12 \pm 1.34$ &     ... \\
     243 &    0 & $346.72821$ & $12.83008$ & $255$ & $20.14$ & $  -9.01$ &  	...  &     ...  & 1 & 23 &   ... & $  -9.01 \pm 1.50$ &     ... \\
      97 &    0 & $346.76483$ & $12.73361$ & $316$ & $19.14$ & $  -5.21$ & $  -4.63$ &     ...  & 2 & 23 &  0.41 & $  -4.92 \pm 0.41$ &     ... \\
      75 &    0 & $346.73433$ & $12.79969$ & $201$ & $18.64$ & $  -3.04$ & $  -2.92$ &     ...  & 2 & 33 &  0.08 & $  -2.98 \pm 0.08$ &     ... \\
     158 &    0 & $346.76642$ & $12.80003$ & $308$ & $19.53$ & $  -1.00$ & $  -0.81$ &     ...  & 2 & 23 &  0.13 & $  -0.91 \pm 0.13$ &     ... \\
      53 &    0 & $346.71908$ & $12.69136$ & $308$ & $18.14$ & $   0.10$ &  	...  & $ -1.10$ & 2 & 23 &  0.86 & $  -0.50 \pm 0.86$ &     ... \\
      87 &    0 & $346.75204$ & $12.81072$ & $275$ & $18.85$ &      ...  & $  -0.03$ &     ...  & 1 & 13 &   ... & $  -0.03 \pm 0.97$ &     ... \\
   28--6 & 0--0 & $346.69946$ & $12.73983$ & $123$ & $17.02$ & $   0.69$ & $   1.32$ &     ...  & 2 & 11 &  0.44 & $   1.01 \pm 0.44$ & $  0.68$\\
     207 &    0 & $346.68692$ & $12.78711$ & $ 53$ & $19.91$ & $   1.49$ & $   2.70$ &     ...  & 2 & 24 &  0.86 & $   2.10 \pm 0.86$ &     ... \\
     262 &    0 & $346.75937$ & $12.70397$ & $357$ & $20.17$ & $   4.48$ & $   3.71$ & $  4.66$ & 3 & 23 &  0.50 & $   4.28 \pm 0.50$ &     ... \\
     198 &    0 & $346.64487$ & $12.74894$ & $166$ & $19.91$ &      ...  & $   3.92$ & $  5.18$ & 2 & 23 &  0.88 & $   4.55 \pm 0.88$ &     ... \\
     148 &    0 & $346.61904$ & $12.72508$ & $290$ & $19.86$ &      ...  & $   5.15$ & $  4.40$ & 2 & 44 &  0.53 & $   4.78 \pm 0.53$ &     ... \\
     238 &    0 & $346.74342$ & $12.70831$ & $306$ & $20.08$ & $   5.19$ & $   5.96$ &     ...  & 2 & 33 &  0.55 & $   5.58 \pm 0.55$ &     ... \\
     217 &    0 & $346.63392$ & $12.73411$ & $227$ & $19.97$ &      ...  & $   6.05$ &     ...  & 1 & 13 &   ... & $   6.05 \pm 1.71$ &     ... \\
     154 &    0 & $346.70625$ & $12.73928$ & $137$ & $19.71$ &      ...  & $   7.07$ &     ...  & 1 & 23 &   ... & $   7.07 \pm 1.25$ &     ... \\
      85 &    0 & $346.60979$ & $12.83125$ & $342$ & $18.80$ & $   7.83$ &  	...  & $  7.61$ & 2 & 12 &  0.16 & $   7.72 \pm 0.16$ &     ... \\
 156--23 & 0--0 & $346.68996$ & $12.75047$ & $ 77$ & $19.50$ & $   6.30$ & $   9.28$ &     ...  & 2 & 13 &  2.11 & $   7.79 \pm 2.11$ & $ 18.57$\\
      36 &    0 & $346.70350$ & $12.68986$ & $295$ & $17.64$ & $   5.65$ & $  10.15$ &     ...  & 2 & 12 &  3.19 & $   7.90 \pm 3.19$ &     ... \\
      60 &    0 & $346.69912$ & $12.82744$ & $200$ & $18.44$ & $   9.58$ & $  10.55$ &     ...  & 2 & 12 &  0.68 & $  10.06 \pm 0.68$ &     ... \\
      23 &    0 & $346.68304$ & $12.70192$ & $246$ & $16.97$ & $  11.71$ & $  12.03$ & $ 11.19$ & 3 & 11 &  0.42 & $  11.64 \pm 0.42$ &     ... \\
\hline		  	    
      38 &    1 & $346.67312$ & $12.78292$ & $ 58$ & $17.81$ & $  35.87$ & $   6.92$ &     ...  & 2 & 12 & 20.47 & $  21.39 \pm20.47$ &     ... \\
112--910 & 1--1 & $346.66404$ & $12.79308$ & $106$ & $19.28$ & $  24.40$ &  	...  & $ 24.33$ & 2 & 24 &  0.05 & $  24.36 \pm 0.05$ & $ 23.77$\\
      22 &    1 & $346.65546$ & $12.82411$ & $212$ & $16.90$ & $  24.80$ & $  24.84$ & $ 23.60$ & 3 & 12 &  0.71 & $  24.41 \pm 0.71$ &     ... \\
 24--118 & 1--0 & $346.70879$ & $12.78747$ & $100$ & $17.00$ & $  25.12$ & $  25.66$ &     ...  & 2 & 11 &  0.38 & $  25.39 \pm 0.38$ & $ 24.92$\\
  41--31 & 1--1 & $346.70812$ & $12.75797$ & $ 95$ & $17.80$ & $  25.70$ & $  25.82$ &     ...  & 2 & 11 &  0.08 & $  25.76 \pm 0.08$ & $ 25.09$\\
      79 &    1 & $346.68458$ & $12.76372$ & $ 29$ & $19.02$ & $  25.85$ &  	...  & $ 25.67$ & 2 & 13 &  0.13 & $  25.76 \pm 0.13$ &     ... \\
 188--96 & 1--1 & $346.69067$ & $12.77817$ & $ 28$ & $19.83$ & $  25.96$ &  	...  &     ...  & 1 & 23 &   ... & $  25.96 \pm 1.01$ & $ 25.38$\\
  37--72 & 1--1 & $346.70212$ & $12.77217$ & $ 60$ & $17.62$ & $  26.59$ & $  26.05$ &     ...  & 2 & 11 &  0.38 & $  26.32 \pm 0.38$ & $ 28.79$\\
     127 &    1 & $346.63037$ & $12.72667$ & $253$ & $19.30$ & $  26.44$ &  	...  &     ...  & 1 & 13 &   ... & $  26.44 \pm 0.42$ &     ... \\
  66--41 & 1--1 & $346.70117$ & $12.76311$ & $ 64$ & $18.58$ & $  26.69$ &  	...  &     ...  & 1 & 12 &   ... & $  26.69 \pm 0.46$ & $ 19.24$\\
  88--36 & 1--1 & $346.68258$ & $12.76011$ & $ 42$ & $18.97$ & $  27.97$ &  	...  &     ...  & 1 & 13 &   ... & $  27.97 \pm 0.54$ & $ 25.29$\\
\hline
\end{tabular}
\label{tableallstars}
\end{table*}

\section{Memberships}

\subsection{Membership from radial velocities}

A  first glance  at Column  (13) of  Table~\ref{tableallstars} reveals
immediately  a clump of  11 stars  whose radial  velocities accumulate
around a velocity close to the systemic radial velocity of Palomar~13,
namely $24.1 \pm 0.5$~km\thinspace s$^{-1}$,  as measured by C\^ot\'e et al.
This clustering of radial velocities  is significant: the $\ensuremath{V_\mathrm{r}}$ of all
the other  stars in Table~\ref{tableallstars} are away  from the above
systemic radial velocity by more  than 10 times the standard deviation
of the $\ensuremath{V_\mathrm{r}}$ of these 11 stars.

These 11 stars are potential cluster members, but before computing any
new estimates of the  systemic radial velocity and projected intrinsic
velocity dispersion  of Palomar~13, we first  compare our measurements
with  the Keck radial  velocities and  check for  possible non-members
using proper motion and photometric data.

\subsection{Agreement with Keck radial velocities}

The  last two columns  in Table~\ref{tableallstars}  provide, whenever
available,   i.e. for  13  stars,  the  radial  velocities  from  both  FLAMES
($\ensuremath{V_\mathrm{r}}$FLAMES)
and Keck ($\ensuremath{V_\mathrm{r}}$Keck) studies.  The  global agreement is  excellent.  From  the 13
stars, 9  lie within the limits  $|\ensuremath{V_\mathrm{r}}\mathrm{FLAMES} - \ensuremath{V_\mathrm{r}}\mathrm{Keck}| <
1.2$~km\thinspace s$^{-1}$ with an average difference of $0.29$~km\thinspace s$^{-1}$.

From the  six  stars with both FLAMES and Keck measurements, which are clearly  non-member  (above  the  intermediate
horizontal line in Table~\ref{tableallstars}), only one, with FLAMES ID
\#156 (Keck ID  \#23), has a radial  velocity difference between two
FLAMES epochs which is larger than the  above $1.2$~km\thinspace s$^{-1}$ limit. This
star, clearly  detected as  variable through our  internal consistency
check, has  a mean $\ensuremath{V_\mathrm{r}}\mathrm{FLAMES} = 7.79 \pm 2.11$~km\thinspace s$^{-1}$ (mean of two
measurements) and  a $\ensuremath{V_\mathrm{r}}\mathrm{Keck} = 18.57 \pm 0.82$~km\thinspace s$^{-1}$ (one
measurement),  with $|\Delta\ensuremath{V_\mathrm{r}}| =  12.27$~km\thinspace s$^{-1}$.   In both
studies, this star is not considered as a cluster member, neither from
both  proper motion nor  from Color-Magnitude-Diagram  (CMD) criteria.
We consider this star as variable according to radial velocity (either pulsating or binary).

The situation  is slightly different  for the seven  potential cluster
members    (below     the    intermediate    horizontal     line    in
Table~\ref{tableallstars}) having both FLAMES and Keck measurements.
\begin{itemize}

\item The star with FLAMES ID \#66  (Keck ID \#41) has a $\ensuremath{V_\mathrm{r}}\mathrm{FLAMES} =
26.69 \pm 0.5$~km\thinspace s$^{-1}$ and  a $\ensuremath{V_\mathrm{r}}\mathrm{Keck} = 19.24 \pm 0.47$~km\thinspace s$^{-1}$, with
$|\Delta\ensuremath{V_\mathrm{r}}|  = 7.45$~km\thinspace s$^{-1}$.  In both studies,  this star is
considered as  a cluster member.  It  is interesting to  note that the
above Keck value is the  mean of three individual measurements, spread
over more than a year, with $\ensuremath{V_\mathrm{r}}\mathrm{Keck} = 18.91 \pm 0.93$~km\thinspace s$^{-1}$, $18.37
\pm 1.84$~km\thinspace s$^{-1}$,  and  $19.43 \pm 0.54$~km\thinspace s$^{-1}$,   in  chronological
order. They do not show any significant variation.

\item There are two more  stars with marginal disagreements: first, stars
with FLAMES ID \#37 (Keck ID  \#72) has a $\ensuremath{V_\mathrm{r}}\mathrm{FLAMES} = 26.32 \pm
0.2$~km\thinspace s$^{-1}$ and a $\ensuremath{V_\mathrm{r}}\mathrm{Keck} = 28.79 \pm 0.27$~km\thinspace s$^{-1}$, with $|\Delta\ensuremath{V_\mathrm{r}}| = 2.47$~km\thinspace s$^{-1}$;
second,  star with FLAMES ID \#88 (Keck ID \#36) has a  $\ensuremath{V_\mathrm{r}}\mathrm{FLAMES} =  27.97 \pm 0.5$~km\thinspace s$^{-1}$\ and a  $\ensuremath{V_\mathrm{r}}\mathrm{Keck} =
25.29 \pm 0.89$~km\thinspace s$^{-1}$, with $|\Delta\ensuremath{V_\mathrm{r}}|  = 2.68$~km\thinspace s$^{-1}$.

\end{itemize}

Are these three  stars pulsating or binaries~?  Two  of them have only
one FLAMES measurement, while the third star, with FLAMES ID \#37 has
two  FLAMES $\ensuremath{V_\mathrm{r}}$  measurements  differing by  $0.54$~km\thinspace s$^{-1}$.  We  discuss
below the influence of these stars on the velocity dispersion estimate
of the cluster.

\subsection{One RR-Lyrae star in the cluster}

The star with FLAMES ID  \#38 has two measurements, with $\ensuremath{V_\mathrm{r}}\mathrm{FLAMES}
= 35.87 \pm 0.70$~km\thinspace s$^{-1}$\ and $6.92 \pm 1.10$~km\thinspace s$^{-1}$, with $|\Delta\ensuremath{V_\mathrm{r}}| = 28.95$~km\thinspace s$^{-1}$. 
 This is a known RR-Lyrae variable \citep[\#\#3 in ][]{Cii_65}, however we ignore it in the remaining of this study in
spite of the fact that it is considered as a cluster member.

\subsection{Membership from stellar proper motions and photometry}

Table~\ref{tablemembership} gives  for the clump  of 11 stars  (at the
bottom  of Table~\ref{tableallstars}), all  of them  potential cluster
members through  their $\ensuremath{V_\mathrm{r}}$,  the following information 
\citep[from][]{Sil_01}.   Columns (1), (2), and  (3) give the ID  number and the
equatorial coordinates  $\alpha$ and  $\delta$. Columns (4),  (5), and
(6) provide the proper motion values $\mu$($\alpha$), $\mu$($\delta$),
and their related membership probability  in \%.  Column (7) gives the
FLAMES radial velocity.  Columns (8), (9) and (10) display the $U,B,V$
photometry, while  Column (11) indicates  the position of the  star on
the CMD:  RRL for RR-Lyrae  pulsating star, RGB for  Red-Giant Branch,
SGB  for Sub-Giant  Branch,  
BS for  Blue Straggler,  and  off-r for  being  off to  the  right  of the  cluster
sequences.

The membership probabilities as determined by \citet{Sil_01} are based on the proper
motion  relative  to  the  average  proper  motion  of the predetermined cluster members. The
probability   is   computed   using,   for   each   individual  candidate  star,  the  ratio
between  the  relative  proper  motion $\mu$ and the measurement error $\sigma_{\mu}$. The
process  is  iterated till a clean sample of cluster members covering the area of study
is obtained.

The      proper      motion      membership      probabilities      in
Table~\ref{tablemembership} give 8 stars out of 11 as clear members of
the cluster, while the 2 stars with  FLAMES ID \#24 and \#127 have
zero  probability of being  members.  The  RR-Lyrae, FLAMES  ID \#38,
with a low membership probability of 48~\% is ignored anyway.

The  CMD  position  in  
Table~\ref{tablemembership} gives 8 stars out of 11 as clear members of
  the cluster, corresponding to the cluster RGB, SGB and BS sequences. Again, the  same  two  stars  with FLAMES ID \#24 and \#127 are off the cluster sequences, to the right of the CMD, although less significantly for the former than the latter. This confirms the selection based on the proper motion.

The star with FLAMES  ID \#79 (not measured in the  Keck study) has a
CMD position  indicating that  it may be  a blue straggler;  given its
$\ensuremath{V_\mathrm{r}}\mathrm{FLAMES} = 25.76 \pm 0.65$~km\thinspace s$^{-1}$\ and its proper motion membership
probability of 65~\%, we consider it as a cluster member.

\subsection{Eight members of Palomar~13~?}

The above discussion illustrates  the danger of considering only radial
velocities in  order to  select cluster members.   The two  stars with
FLAMES ID \#24 and \#127 (the latter not measured in the Keck study)
with  $\ensuremath{V_\mathrm{r}}\mathrm{FLAMES} = 26.44 \pm  0.42$~km\thinspace s$^{-1}$ and $25.39 \pm 0.21$~km\thinspace s$^{-1}$
respectively,  have $\ensuremath{V_\mathrm{r}}$ values  very close  to the  cluster systemic
radial velocity.   However, both proper motions  probabilities and CMD
positions make us consider them as non-members of the cluster. 

From the examples above, it becomes  obvious that to declare a star as
a  member or not of  Palomar~13 represents a  difficult task, which 
needs the  compulsory simultaneous  use of at  least the  three sorting 
processes available  so far, viz radial  velocities, proper motions, 
and positions in the CMD. 

We  are  left  with  a  sample  of  eight  stars  probable  member  of 
Palomar~13.

\begin{table*}[!htp]
\caption{Membership probabilities  from proper motions  and positions in 
the CMD for the 11 stars with radial velocities implying membership of 
Palomar~13.} 
\setlength{\tabcolsep}{2.5pt}
\begin{tabular}{cccrrcccccc}
\hline 
\hline 
\multicolumn{1}{c}{ID}  & \multicolumn{1}{c}{$\alpha_\mathrm{J2000}$}
& \multicolumn{1}{c}{$\delta_\mathrm{J2000}$}  &
\multicolumn{1}{c}{$\mu$($\alpha$)} &
\multicolumn{1}{c}{$\mu$($\delta$)} & \multicolumn{1}{c}{P($\mu$)}  &
\multicolumn{1}{r}{\ensuremath{V_\mathrm{r}} FLAMES}    & \multicolumn{1}{c}{$U$}   & \multicolumn{1}{c}{$B$}   & \multicolumn{1}{c}{$V$}   & \multicolumn{1}{c}{CMD} \\
           & \multicolumn{1}{c}{[$^\mathrm{h\;\;m\;\;s}$]} & 
	     \multicolumn{1}{c}{[$^\circ\; ^\prime\; ^{\prime\prime}$]} & 
	     \multicolumn{1}{c}{[mas\,yr$^{-1}$]}          & 
	     \multicolumn{1}{c}{[mas\,yr$^{-1}$]}          & 
	     \multicolumn{1}{c}{[\%]}        & 
	     \multicolumn{1}{c}{[km\thinspace s$^{-1}$]}      & 
	     \multicolumn{1}{c}{[mag]}   & \multicolumn{1}{c}{[mag]}   & \multicolumn{1}{c}{[mag]}  \\
\multicolumn{1}{c}{(1)} & \multicolumn{1}{c}{(2)}       & \multicolumn{1}{c}{(3)}       & \multicolumn{1}{c}{(4)}             & \multicolumn{1}{c}{(5)}             & \multicolumn{1}{c}{(6)}       & \multicolumn{1}{c}{(7)}      & \multicolumn{1}{c}{(8)}   & \multicolumn{1}{c}{(9)}   & \multicolumn{1}{c}{(10)}  & \multicolumn{1}{c}{(11)} \\
\hline
  38 & 23 06 41.55 & 12 46 58.5 & $ 0.32 \pm 0.11$ & $ 0.05 \pm 0.19$ & 48 & 21.39 & $18.432 \pm 0.020$ & $18.335 \pm 0.010$ & $17.808 \pm 0.007$ &  RRL   \\
 112 & 23 06 39.37 & 12 47 35.1 & $-0.31 \pm 0.16$ & $ 0.59 \pm 0.29$ & 89 & 24.36 & $20.134 \pm 0.042$ & $20.070 \pm 0.021$ & $19.276 \pm 0.014$ &  SGB   \\
  22 & 23 06 37.31 & 12 49 26.8 & $-0.22 \pm 0.07$ & $-0.02 \pm 0.10$ & 96 & 24.82 & $18.148 \pm 0.023$ & $17.761 \pm 0.012$ & $16.897 \pm 0.007$ &  RGB   \\
  24 & 23 06 50.11 & 12 47 14.9 & $ 0.24 \pm 0.19$ & $ 1.6  \pm 0.16$ &  0 & 25.39 & $18.334 \pm 0.020$ & $17.983 \pm 0.011$ & $17.005 \pm 0.007$ &  off-r?\\
  41 & 23 06 49.95 & 12 45 28.7 & $-0.05 \pm 0.12$ & $-0.25 \pm 0.12$ & 79 & 25.76 & $18.854 \pm 0.022$ & $18.658 \pm 0.011$ & $17.795 \pm 0.007$ &  RGB   \\
  79 & 23 06 44.30 & 12 45 49.4 & $ 0.22 \pm 0.16$ & $-0.36 \pm 0.18$ & 65 & 25.76 & $19.418 \pm 0.027$ & $19.386 \pm 0.014$ & $19.021 \pm 0.012$ &  BS?   \\
 188 & 23 06 45.76 & 12 46 41.4 & $-0.18 \pm 0.20$ & $ 0.09 \pm 0.23$ & 99 & 25.96 & $20.535 \pm 0.058$ & $20.529 \pm 0.024$ & $19.834 \pm 0.018$ &  SGB   \\
  37 & 23 06 48.51 & 12 46 19.8 & $ 0.00 \pm 0.09$ & $ 0.32 \pm 0.09$ & 88 & 26.32 & $18.757 \pm 0.023$ & $18.537 \pm 0.011$ & $17.625 \pm 0.007$ &  RGB   \\
 127 & 23 06 31.29 & 12 43 36.0 & $-4.4  \pm 0.77$ & $-6.2  \pm 0.54$ &  0 & 26.44 & $21.740 \pm 0.185$ & $20.579 \pm 0.034$ & $19.305 \pm 0.018$ &  off-r \\
  66 & 23 06 48.28 & 12 45 47.2 & $-0.41 \pm 0.13$ & $ 0.03 \pm 0.14$ & 88 & 26.69 & $19.427 \pm 0.028$ & $19.361 \pm 0.014$ & $18.583 \pm 0.010$ &  SGB   \\
  88 & 23 06 43.82 & 12 45 36.4 & $-0.30 \pm 0.18$ & $-0.20 \pm 0.20$ & 93 & 27.97 & $19.842 \pm 0.033$ & $19.727 \pm 0.016$ & $18.970 \pm 0.011$ &  SGB   \\
\hline
\end{tabular}
\label{tablemembership}
\end{table*}

\section{Velocity dispersion and mass-to-light ratio}

Given the limited size of our sample which does not permit any attempt
to discuss  any concentric variation  of any kinematic  quantity, we
focus here on  the best estimate of the  unbiased mean radial velocity
dispersion  for  the  whole   sample.   The  crucial  element  is  the
subtraction of the  random error of our sample  and the error estimate
of our  velocity dispersion.  We  compute the velocity  dispersion and
the  mass-to-light ratio  in four  different ways:  (i) for  the eight
stars with  at least one FLAMES  $\ensuremath{V_\mathrm{r}}$ value, (ii) for  the five stars
with two  FLAMES $\ensuremath{V_\mathrm{r}}$ values (measured  twice), and then  for the six
stars which  are measured in both  Keck and FLAMES  studies, (iii) for
the six FLAMES $\ensuremath{V_\mathrm{r}}$ values and (iv) for the six Keck $\ensuremath{V_\mathrm{r}}$ values.

The average empirical variance of  radial velocity of 31 stars with at
least two measurements  is $var_0 = 0.29$~km$^2$\,s$^{-2}$.  This  is the most
realistic  estimate  of  the   measurement  error  applicable  to  all
measurements.   The   stars  with two  measurements  are   taken  with
$var=0.5\,var_0$ while stars with  only one measurement with $var=var_0$.
We  compute  the unbiased  systemic  cluster  velocity $\langle\ensuremath{V_\mathrm{r}}\rangle$,   
the  intrinsic   projected  cluster   velocity  dispersion
$\sigma_\mathrm{p}$ and associated errors  according to the method described in
\citet{PrrMen93}, which were also used in the Keck study (C\^ot\'e et al.).

\subsection{Sample of eight stars with at least one FLAMES $\ensuremath{V_\mathrm{r}}$ value }

The systemic  velocity of  Palomar~13 from our  eight member  stars is
$\langle V_\mathrm{r8}\rangle = 25.91 \pm 0.36$~km\thinspace s$^{-1}$ with a projected velocity
dispersion $\sigma_\mathrm{p8}=0.92 \pm 0.29$~km\thinspace s$^{-1}$.  This $\sigma_\mathrm{p8}$ value
should be considered as an upper  limit to the real value since, for a
given sample, any source of unknown random error will tend to increase
the  measured velocity  dispersion. The  above value  is significantly
smaller than $\sigma_\mathrm{p}  = 2.2 \pm 0.4$~km\thinspace s$^{-1}$ found  by C\^ot\'e et al.

Note that  even a  very crude unweighted  computation of  the systemic
velocity   and  projected   velocity  dispersion   will   not  produce
significantly  different results:  namely, for  the same  eight stars,
$\langle\ensuremath{V_\mathrm{r}}\rangle =  25.95$~km\thinspace s$^{-1}$  and $\sigma_\mathrm{p}=1.04$~km\thinspace s$^{-1}$.   In
general terms, we  can say that whatever  the sophistication of the
weighting  process  used  in  the  above  computation,  the  resulting
velocity dispersion will always be  close to $1$~km\thinspace s$^{-1}$.  This is a robust
result.

What  is the mass-to-light  ratio corresponding  to this  new velocity
dispersion $\sigma_\mathrm{p}$  value~?  If only the  $\sigma_\mathrm{p}$ value changes,
everything  else  being equal  (King-Michie  model,  core radius,  and
central surface  brightness, see Table~3 from C\^ot\'e et al.),
the value of the $\mathcal{M}/L_V$ ratio varies as the squared ratio
of  the  old  to  new  velocity dispersions.   Consequently,  the  new
$\mathcal{M}/L_V$ is equal to the old $\mathcal{M}/L_V$ divided by
$(2.2/0.92)^2 = 5.8$, namely, $\mathcal{M}/L_V =  40/5.8 \sim 7 $,
a value  significantly closer  to the usual  value $\mathcal{M}/L_V \sim 3$ for globular clusters.

Since  we  have  a  rather   small  sample,  we  explore  briefly  the
sensitivity of the above results to various subsamples.

\subsection{Sample of five stars with  two FLAMES  $\ensuremath{V_\mathrm{r}}$ measurements}

Similar arithmetics applied  to the five Palomar~13 member stars with at
least two FLAMES measurements gives the values $\langle V_\mathrm{r5}\rangle = 25.40 \pm 0.32$~km\thinspace s$^{-1}$
and $\sigma_\mathrm{p5}=0.60\pm 0.27$~km\thinspace s$^{-1}$.   This again may be  considered as an
upper  limit to the  real value.   From $(2.2/0.60)^2  = 13.4$,  the new
$\mathcal{M}/L_V =  40/13.4  \sim 3.0 $,  a value identical  to the
usual value for globular clusters.

\subsection{Sample of six stars with FLAMES and Keck $\ensuremath{V_\mathrm{r}}$ measurements}

Six of the  above eight Palomar~13 stars have  $\ensuremath{V_\mathrm{r}}$ measurements from
both FLAMES and Keck studies.

The six  $\ensuremath{V_\mathrm{r}}$ measurements  from this FLAMES  study bring  the values
$\langle V_\mathrm{r6F}\rangle = 26.14 \pm 0.44$~km\thinspace s$^{-1}$  and  $\sigma_\mathrm{p6F}=0.99 \pm 0.34$~km\thinspace s$^{-1}$.   The
corresponding $\mathcal{M}/L_V$ would scale by $(2.2/0.99)^2 = 4.9$,
namely, $\mathcal{M}/L_V = 40/4.9 \sim 8 $.

The  six $\ensuremath{V_\mathrm{r}}$  measurements  from  the Keck  study  bring the  values
$\langle V_\mathrm{r6K}\rangle  =  24.60 \pm 1.16$~km\thinspace s$^{-1}$ and
$\sigma_\mathrm{p6K}=2.79 \pm 0.83$~km\thinspace s$^{-1}$.   The
corresponding  $\mathcal{M}/L_V$  would  scale by  $(2.2/2.79)^2=0.62$,
 namely, $\mathcal{M}/L_V = 40/0.62 \sim 60 $.

For the  same stars,  FLAMES $\ensuremath{V_\mathrm{r}}$ values  produce a  smaller velocity
dispersion than the Keck $\ensuremath{V_\mathrm{r}}$ values, indication of a possible better
stability  and  better  zero-point  calibration for  the  simultaneous
FLAMES measurements than for the independent Keck ones.

\subsection{Overlap of Keck and FLAMES sample fields}

Because  of  the positioning  of  FLAMES  fibres,  which prevents  the
measurements of stars with small separations, the star sample acquired
during this commissioning  observing run is not optimal  for the study
of Palomar~13.  We do not  have numerous member stars within the tidal
radius  of  the cluster  as  determined  by C\^ot\'e et al.
Following our three criteria (radial velocities, proper motions, and CMD
photometry), there seems  to be members of Palomar~13  well outside the
nominal tidal radius, as already noticed by C\^ot\'e et al.

Could  our  small  velocity dispersion value originate from the fact that we have stars on
average  more  distant  from the cluster center than those in the Keck study~? This is not
the case. We do not observe any central increase in velocity dispersion in Keck data since
the  stars  contributing  strongly  to  the velocity dispersion lie all outside the radius
60~\arcsec.
 


\section{Final remarks}

Although small, the present  sample of radial velocities measured with
FLAMES provides velocity  dispersion and mass-to-light ratio estimates
which  differ  significantly  from  those obtained  from  Keck  radial
velocities.

In  both,  FLAMES  and  Keck  samples, the detection of binary stars is difficult.
C\^ot\'e et al.  have good time coverage (up to four epoch) but few stars measured
more than once, while FLAMES has two measurements for most stars but no time coverage. Two
stars, \#66-41 and \#37-72 are responsible for significant differences between FLAMES and Keck
results.  Though  these objects have not shown significant variability in, respectively, three
and four Keck measurements and fall in the $\ensuremath{V_\mathrm{r}}$ range of the cluster in FLAMES data, we must
accept  them  as  binary  candidates. Based on the proportion of blue stragglers in CMD of
Palomar~13,  approximately  25\%  of evolved stars are binary systems, a figure well compatible
with two binaries found and simulations carried out by C\^ot\'e  et~al.

 The difference between FLAMES and Keck results 
casts  doubt  on  the  previous  Keck velocity dispersion determination. The most simple and
likely  explanation  of  the  very  high Keck $\mathcal{M}/L_V$ ratio is the Keck sample
contamination  by  binaries  since  the determination of velocity dispersion even from the
larger  (still  small)  sample  from  Keck  is  prone  to be biased by the presence of few
undected binaries.

A  complete  and  more   detailed  analysis  of  these  FLAMES  radial
velocities along with the old and some new radial velocities from Keck
is under preparation, with extensive simulations of such observations.
A careful  attention will be given  to the surface  density profile of
Palomar~13  which  anyway  appears  to  be  anomalous  among  Galactic
globular clusters.
Is there  anything  genuinely peculiar  with  Palomar~13~?  A  significant
effort (new  observations and  new analysis) is  needed to  answer the
following  question: Is  Palomar~13 the  host of  a large  fraction of
binaries  which  may inflate  its velocity  dispersion~?
With  a  good  observing   strategy,  FLAMES  is  currently  the  best
instrument to answer this question.


\begin{acknowledgements}
      We  thank  Pat~C\^ot\'e for the image used in Fig. 2 as finding chart.
      We are very grateful to G.~Simond for his work in the development of the GIRAFFE Data Reduction Software, G.~Burki
      and  M.~Grenon  for usefull discussions and the FLAMES commissioning team for the data.
      Part  of  this  work  was  supported  by  Swiss  {\em  Fond National de la Recherche
      Scientifique.}

\end{acknowledgements}



\end{document}